\documentclass[aps,pra,twocolumn,showpacs]{revtex4}

\usepackage[dvips]{graphicx}
\usepackage[normalem]{ulem}
\usepackage{graphicx}
\usepackage{dcolumn}
\usepackage{bm}
\input{epsf}
\usepackage{color}

\newcommand{\be}{\begin{equation}}
\newcommand{\ee}{\end{equation}}
\newcommand{\beq}{\begin{eqnarray}}
\newcommand{\enq}{\end{eqnarray}}
\newcommand{\ua}{\uparrow}
\newcommand{\da}{\downarrow}
\newcommand{\uden}{n_\uparrow}
\newcommand{\dden}{n_\downarrow}
\newcommand{\nup}{N_\uparrow}
\newcommand{\ndo}{N_\downarrow}
\newcommand{\eup}{E_{F\uparrow}}
\newcommand{\ox}{\omega_x}
\newcommand{\oy}{\omega_y}
\newcommand{\oz}{\omega_z}
\newcommand{\Om}{\Omega}
\newcommand{\op}{\omega_\perp}
\newcommand{\rsf}{R_{\rm S}(\theta,\phi)}
\newcommand{\suf}{R_{\rm S}}
\newcommand{\sun}{R_{\rm N}}
\newcommand{\rn}{R_{\rm N}(\theta,\phi)}

\newcommand{\rms}{\rm S}
\newcommand{\rmn}{\rm N}
\newcommand{\eps}{\epsilon}

\newcommand{\pc}{P_{\rm c}}
\newcommand{\xc}{x_{\rm c}}

\begin{document}


\title{Unitary polarized Fermi gas under adiabatic rotation}

\author{I. Bausmerth$^{1}$}\email{ingrid@science.unitn.it}
\author{A. Recati$^{1,2}$}
\author{S. Stringari$^{1}$}
\affiliation{$^{1}$Dipartimento di Fisica, Universit\`a di Trento and CRS-BEC INFM,
I-38050 Povo, Italy\\
$^{2}$Physik Department, Technische Universit\"at M\"unchen, D-85747 Garching, Germany} 

\begin{abstract}

We discuss the effect of an adiabatic rotation on  the phase separation between the superfluid and  normal component of a trapped polarized Fermi gas at unitarity and zero temperature, under the assumption that quantized vortices are not formed. We show that the Chandrasekhar-Clogston limit $\dden/\uden$ characterizing the local polarization in the normal phase at the interface is enhanced by the rotation as a consequence of the centrifugal effect. The density profiles (local and column integral) of the two spin species are calculated as a function of the angular velocity for different values of the polarization. The critical value of the angular velocity at which the superfluid exhibits a spontaneous quadrupole deformation is also calculated for the unpolarized case. 

\end{abstract}


\maketitle

\section{Introduction}

Recent experiments \cite{MIT1,MIT2,MIT3,Rice,Rice2} have shown that a polarized Fermi gas at unitarity and zero temperature undergoes a phase separation between a central unpolarized superfluid component and an external polarized normal gas. 
Experiments, where surface tension effects are not important and the local density approximation (LDA) is applicable, have revealed the occurrence of a critical value of the polarization
\begin{equation}
P=\frac{\nup-\ndo}{\nup+\ndo}
\label{P}
\end{equation}
of the gas above which the superfluid component disappears. It is found that such a critical value is $\pc\simeq77\%$.
 
The study of spin-polarized Fermi gases has been the object of several theoretical papers. In \cite{normalLobo,recati} the equations of state of uniform matter for the superfluid and for the polarized normal phases, calculated with {\it ab initio} Monte Carlo simulations, have been employed within a local density approximation to treat the effect of the harmonic trapping in the unitary regime \cite{bulnote}. These calculations were proven to be very efficient not only in reproducing the experimental value of the critical polarization, but also the density profiles of the two separate spin components. Therefore they provide an accurate and consistent description of the phase separation  exhibited by the unitary Fermi gas at zero temperature. In particular, the  discontinuity characterizing the spin-down density at the interface between the superfluid and normal components,  as well as the typical knee revealed by the column density of the same spin-down component, are dramatic features reproduced with high accuracy by theory. These calculations have pointed out the crucial role played by the interactions in the normal phase \cite{recati}. 

\begin{figure}[htb]
\begin{center}
\includegraphics[height=5cm] {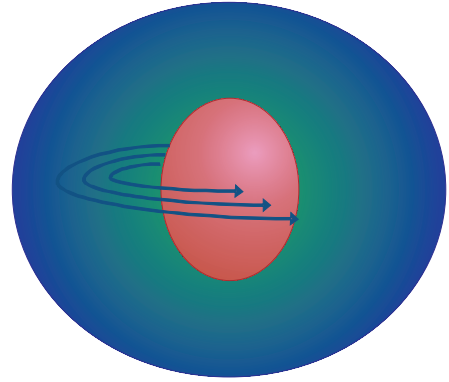}
\end{center}
\caption{(Color online) The typical shell structure of the trapped system  consisting of a  superfluid core (red) surrounded by a partially (green) and fully polarized (blue) normal shell for a polarization $P=44\%$ and $\Om=0.5\op$. The superfluid is squeezed in the radial direction while the normal part exhibits the bulge effect of the rotation.}
\label{fig:sketch}
\end{figure}

The purpose of this paper is to explore the rotational properties of these polarized Fermi gases (see the sketch in Fig. \ref{fig:sketch}). It is well known that the response to a transverse probe, like the rotation, is a crucial tool to test the superfluidity of a system. In a trap, this can be achieved by rotating the confining potential. While a normal gas rotates in a classical rigid way, a superfluid features a different behavior.  The phenomena exhibited by  superfluids include the quenching of the moment of inertia at small angular velocities and  the appearance of quantized vortices at higher velocities. Vortices have already been observed in these polarized configurations and shown to disappear for high polarizations. In a recent  paper \cite{ias} we have predicted that a trapped rotating  Fermi gas at unitarity exhibits a further interesting phenomenon, associated with the breaking of superfluidity in the external region. Indeed, atoms prefer to be in the normal phase because of the energy gain associated with the rotation. This mechanism of the depletion of the superfluid due to the rotation has also been recently  confirmed within BCS mean-field theory \cite{urban}. The occurrence of this  phenomenon requires proper conditions of adiabaticity in the ramping of the rotation of the trap in order to avoid the formation of vortices, a condition that has been already successfully realized in  rotating Bose-Einstein condensates \cite{Dalibard1}. 

In a polarized Fermi gas the phase separation between the superfluid and the normal component is already present in the absence of rotation, so that the effect of the rotation is expected to further enhance the mechanism of depletion of the superfluid with new visible consequences on the density profiles of the two spin components. The explicit investigation of these features represents the main goal of the present paper.

This paper is organized as follows. In Sec. \ref{trap} we introduce our model and investigate the effect of rotation on the phase separation between normal and superfluid within the LDA. The results are discussed in Sec. \ref{results}. In Sec. \ref{symmbreak} we calculate the critical frequency for the breaking of axi-symmetry of the superfluid core and finally in Sec. \ref{con} we draw our conclusions.

\section{Polarized Fermi Gas in a rotating Trap}
\label{trap}

As already mentioned throughout the paper we assume that only two homogeneous phases are possible: A polarized normal phase and an unpolarized superfluid one \cite{noteUrban}. 

The normal state of a polarized Fermi gas at unitarity has been first introduced and discussed in \cite{normalLobo,chevy,bulgac}. Figuratively it can be understood as a sea of spin-$\ua$ atoms to which constantly spin-$\da$ atoms are added. Since at unitarity there are no interaction length scales, the energy of the system at zero temperature can be written in terms of the Fermi momentum $k_{\rm F\ua}$ of the majority component and  the concentration $x={\dden}/{\uden}$ of the minority atoms.
As soon as the concentration reaches the critical value $\xc=0.44$ the system starts to nucleate a superfluid region.

In the unitary limit of infinite scattering length the dependence of the energy of the normal phase on the concentration $x$ can be written as \cite{normalLobo,recati} 
\beq
\frac{E_{\rm N}(x)}{N_\ua}&=&\frac{3}{5}\eup\left(\frac{}{}1-Ax+\frac{m}{m^*}x^{5/3}+B x^2\frac{}{}\right)\nonumber\\
&=&\frac{3}{5}\eup g(x)\equiv\eps_{\rmn}(x),  
\label{eq:energyx}
\enq
where $N_\ua$ is the total number of $\ua$ atoms, $\eup=\hbar^2/2m (6 \pi^2 \uden)^{2/3}$ is the ideal gas Fermi energy, and $m$ the atomic mass. 
The values of the spin-down single-particle energy $A$ and of the effective mass $m^*$ have been calculated both analytically and numerically \cite{normalLobo,Combescot,Proko}, the parameter $B$ is known by fitting Monte Carlo results. 
The most recent Monte Carlo calculations give $A=0.99(1)$, $m^*=1.09(2)$, and $B=0.14$ \cite{sebastiano}. The parametrization (\ref{eq:energyx}) reproduces the Monte Carlo results for the energy of the normal state not only in the low $x$ regime, but also for large values of the concentration parameter. 

On the other hand the equation of state for the superfluid phase is simply given by 
\be
\frac{E_{\rm S}}{N_{\rm S}}=\xi_{\rms}\frac{3}{5}\frac{\hbar^2}{2m}(6\pi^2 n_{\rm S})^{2/3}\equiv\eps_{\rms}(n_{\rms}),
\label{eq:ensf}
\ee
where $N_{\rm S}$ is the number of atoms in the superfluid phase, $n_{\rm S}$ the superfluid density (equal to the spin-up and spin-down density) and the interaction parameter  $\xi_{\rms}=0.42$ is known from {\it ab initio} quantum Monte Carlo simulations  \cite{Carlsonxi,Stefanoxi}.

We consider a polarized Fermi gas at unitarity confined by a harmonic potential
$V({\bf r})=m(\ox^2 x^2+\oy^2 y^2+\oz^2 z^2)/2$, rotating  with angular velocity $\Omega$ along the $z$ axis. We study the problem in the rotating frame of the trap, where the potential is static and the Hamiltonian contains the additional term $-\Omega L_{\rm Z}$.
We stress once again that the response of a superfluid to an external rotation is multifaceted and can depend on whether one ramps up the angular velocity very fast or in an adiabatic way. In the following we will not take into account the formation of vortices. This is best ensured by  assuming that the angular velocity is ramped up adiabatically.

In the  LDA the grand canonical energy of the rotating configuration at zero temperature takes the form
\beq
E&=&\int d{\bf r}\left(\frac{}{}\epsilon(n_{\ua}({\bf r}),n_{\da}({\bf r}))+V({\bf r})\right.\nonumber\\
& &\hspace{0.4cm}+\left.\frac{1}{2}mv^2-m\Omega({\bf r}\times{\bf v})_{\rm Z}\right) n({\bf r})\nonumber\\
& &\hspace{0.4cm}-\int d{\bf r}[\mu_{\ua}^0n_{\ua}({\bf r})+\mu_{\da}^0n_{\da}({\bf r})],
\label{eq:energyPdiff}
\enq
where $\epsilon(n_{\ua}({\bf r}),n_{\da}({\bf r}))$ is the energy density per particle depending on the $n_{\ua,\da}({\bf r})$ densities of the two spin species,  ${\bf v}$ is the velocity field, $\mu_{\ua}^0$ and $\mu_{\da}^0$ are the chemical potentials  of the $\ua$ and $\da$ particles, and  $n({\bf r})=n_{\ua}({\bf r})+n_{\da}({\bf r})$ is the total density. 

We assume that the phase separation in the trap manifests as the formation of an inner unpolarized superfluid core occupying the region $r<\rsf$
surrounded by an external normal shell, which is confined to $\rsf<r<\rn$. Here, we term $\rsf$ the interface separating the superfluid from the normal phase and $\rn$ the Thomas-Fermi radius of the gas where the density vanishes.
Thus, the integral (\ref{eq:energyPdiff}) splits into two parts
\beq
E&=&2 \int_{r<\suf} d{\bf r}\left(\frac{}{}\epsilon_{\rms}(n_{\rms}({\bf r}))-\mu_{\rms}^0\right.\nonumber\\
& &\hspace{0.3cm}\left.+\ V({\bf r})+\frac{1}{2}mv_{\rms}^2-m\Omega({\bf r}\times{\bf v_{\rms}})_{\rm Z}\right) n_{\rms}({\bf r})\nonumber\\
& &\hspace{0.3cm}+\ \int_{\suf<r<\sun} d{\bf r}\ \left[\frac{}{}\eps_{\rmn}(x({\bf r}))n_{\ua}({\bf r})\right.\nonumber\\
& &\left.\hspace{0.3cm}-\ \frac{}{}\mu_{\ua}^0n_{\ua}({\bf r})-\mu_{\da}^0n_{\da}({\bf r})\frac{}{}\right]\nonumber\\
& &\hspace{0.3cm}+\left(\frac{}{} V({\bf r})+\frac{1}{2}mv_{\rmn}^2-m\Omega({\bf r}\times{\bf v_{\rmn}})_{\rm Z}\frac{}{}\right) n({\bf r}),\nonumber\\
\label{eq:energysplit}
\enq
where $\mu_{\rms}^0=(\mu_{\ua}^0+\mu_{\da}^0)/2$ is the superfluid chemical potential and $n_{\ua,\da}({\bf r})$ the  $\ua$ and $\da$ densities in the normal phase. In the above equation we have distinguished between the velocity fields  ${\bf v_{\rms}}$ and ${\bf v_{\rmn}}$ in the superfluid and normal phases, respectively.

To find the equilibrium conditions, we minimize the energy with respect to the densities, to the velocity fields as well as with respect to the border surface.
In the case of the superfluid the velocity field obeys the irrotationality constraint and can thus be written as ${\bf v}_{\rms}=\nabla \Phi$. Variation of the energy with respect to the velocity potential $\Phi$ yields the continuity equation
\be
\nabla\cdot n_{\rms}(\nabla\Phi-{\bf \Omega\times r})=0,
\label{eq:graddens}
\ee
while the variation with respect to the superfluid density $n_{\rms}$ yields the LDA relationship
\be
\mu_{\rm S}^0=\xi_{\rm S}\frac{\hbar^2}{2m}(6\pi^2 n_{\rm S})^{{2}/{3}}+V^{\rms}({\bf r}),
\label{eq:locdensSFPdiff}
\ee
where $V^{\rm S}({\bf r})=V({\bf r})+\frac{1}{2}mv^2-m\Omega({\bf r}\times{\bf v_{\rms}})_{\rm Z}$ is the effective harmonic potential felt by the superfluid.

Using the same procedure for the normal part (without the irrotationality constraint) we get $\bf v_{\rm N}=\Om\times r$, i.e. it rotates rigidly.
The variation with respect to the densities gives the LDA expressions
\be
\mu_{\ua}^0=\left(g(x)-\frac{3}{5}x g'(x)\right)\frac{\hbar^2}{2m}(6\pi^2 n_{\ua})^{{2}/{3}}+V^{\rmn}({\bf r}),
\label{eq:locdensmup}
\ee
\be
\mu_{\da}^0=\frac{3}{5}g'(x)\frac{\hbar^2}{2m}(6\pi^2 n_{\ua})^{{2}/{3}}+V^{\rmn}({\bf r}),
\label{eq:locdensmud}
\ee
where the effective potential $V^{\rmn}({\bf r})$ felt by the particles in the normal phase is now squeezed due to the rigid rotation according to $(\ox^{\rm N})^2=\ox^2-\Omega^2$, $(\oy^{\rm N})^2=\oy^2-\Omega^2$.

By varying the energy (\ref{eq:energysplit}) with respect to $R_{\rm S}(\theta,\phi)$ we eventually find the equilibrium condition for the coexistence of the two phases in the trap.
This is equivalent to implying that the pressure of the two phases be the same
\be
\left(n_{\rms}^2\frac{\partial{\epsilon_{\rms}}}{\partial n_{\rms}}\right)_{r=\suf}=\frac{1}{2}\left(n_{\ua}^2\frac{\partial{\epsilon_{\rmn}(x)}}{\partial n_{\ua}}+n_{\da}n_{\ua}\frac{\partial{\epsilon_{\rmn}(x)}}{\partial n_{\da}}\right)_{r=\suf}.
\label{eq:press}
\ee
Using the expressions for the energy densities (\ref{eq:energyx}) and (\ref{eq:ensf}) we obtain an equation for the density discontinuity in the trap given by
\be
\frac{n_{\ua}(\suf)}{n_{\rms}(\suf)}=\left(\frac{2\xi_{\rms}}{g(x(\suf))}\right)^{3/5}\equiv \gamma(x(\suf)),
\label{eq:jumptrap}
\ee
where $x(\suf)$ is the local concentration at the interface.
Combining the above equation with Eqs.(\ref{eq:locdensSFPdiff}) and (\ref{eq:locdensmup}) we find the useful relationship
\be
\left(\mu_{\rms}^0-V^{\rms}({\suf})\right)=\gamma(x({\suf}))\left(\mu_{\ua}^0-V^{\rmn}({\suf})\right),
\label{eq:defbordergen}
\ee
determining the surface $\rsf$ separating the superfluid and the normal part.

Eventually using also Eq.(\ref{eq:locdensmud}) we find an expression which implicitly defines the concentration as a function of the position at the interface
\beq
g(x({{\suf}}))&+&\frac{3}{5}[1-x({\suf})]g'(x({\suf}))\nonumber\\
&-&(2\xi_{\rms})^{3/5}[g(x({\suf}))]^{2/5}\nonumber\\
&=&\frac{2(V^{\rms}({\suf})-V^{\rmn}({\suf}))}{\eup({\suf})}.
\label{eq:condtrap}
\enq
In absence of rotation $V^{\rm S}=V^{\rm N}\equiv V$ the solution of Eq.(\ref{eq:condtrap}) yields the value $x(R_{\rm S})=0.44$ \cite{normalLobo}. This value coincides with the maximum  concentration achievable in the  normal phase of uniform matter before phase separation. Since the rotation affects differently the potentials $V^{\rm S}$ and $V^{\rm N}$, the value of the concentration depends now on the angular position of the interface. It ranges from a minimum value $x[R_{\rm S}(0,\phi)]=\xc=0.44$ along the $z$ axis (where the effect of the rotation is vanishing) to a maximum, $\Omega$ dependent value $x[R_{\rm S}(\pi/2,\phi)]$ in the $xy$ plane (where the effect of the rotation is largest). The effect of the rotation is thus to enhance the average value of the concentration in the normal phase and hence to favour the depletion of the superfluid (see Fig. {\ref{fig:NS}} and discussion below). This is physically understood by noticing that in the rotating frame the atoms in the normal part gain the energy
$\frac{1}{2}m{\bf v}^2_{\rmn}(\bf r)$ due to the centrifugal force. Thus, the energy of the normal part can become smaller than the value in the superfluid \cite{ias}. The main effect is to change the critical concentration $\xc$ at the interface (see Fig. \ref{fig:conzP44} below).

Notice, however, that the critical {\sl global} concentration $\pc$  for the system to start nucleating the central superfluid core is not affected by the rotation and keeps the nonrotating value $\pc=77\%$ (see also Fig. \ref{fig:conPOL}). The reason for that is easily explained within the local density approximation used here. 
Just above $\pc$ the system is completely normal and the only effect of the rotation is the squeezing of the transverse trapping frequencies, while in the $z$ direction the system remains unaffected (see Eqs.(\ref{eq:locdensmup}, \ref{eq:locdensmud})). Since at $\pc$ the superfluid is nucleated at the center of the trap where centrifugal effects are absent, the local condition for equilibrium between the superfluid and the normal component is the same as without rotation. The calculation of the critical polarization proceeds then in exactly the same way as without rotation, but for a simple rescaling of the trapping frequencies which has no effect on the value of $\pc$.

\section{Results}
\label{results}

In the following we will assume $\ox=\oy=\op$ and we consider the solution $\bf v_{\rm S}=0$, thus a nonrotating axisymmetric superfluid  and consequently $V^{\rm S}\equiv V$. In this case the local concentration depends only on the polar angle $\theta$ and the superfluid radius takes the form
\beq
R_{\rm S}^2(\theta)&=&\frac{2}{m}\frac{(\mu_{\rms}^0-\gamma(x(\suf))\mu_{\ua}^0)}{(1-\gamma(x(\suf)))}\times\nonumber\\
&\ &\left(\frac{}{}\omega_z^2\cos^2\theta+\op^2 \sin^2\theta\right.\nonumber\\
&+&\left.\frac{\gamma(x(\suf))}{1-\gamma(x(\suf))}\Omega^2 \sin^2\theta\right)^{-1},
\label{eq:RSFpolsymm}
\enq 
while the Thomas-Fermi radius $\sun$ of the normal gas is fixed  by the condition $\mu_{\ua}=V^{\rmn}({\bf r})$ yielding
\be
R_{\rm N}^2(\theta)=\frac{2\mu_{\ua}^0}{m}\left(\oz^2\cos^2\theta+\op^2\sin^2\theta-\Omega^2\sin^2\theta\right)^{-1}\geq R_{\rm S}^2(\theta).
\label{eq:RNpolsymm}
\ee
The values of the chemical potentials are fixed by the normalization 
\be
\int_{r<R_{\rm S}}d{\bf r}\  \ n_{\rms}({\bf r})+\int_{R_{\rm S}<r<R_{\rm N}}d{\bf r}\ n_{\ua}({\bf r})=N_{\ua},
\ee
and
\be
\int_{r<R_{\rm S}}d{\bf r}\  \ n_{\rms}({\bf r})+\int_{R_{\rm S}<r<R_{\rm N}}d{\bf r}\ n_{\da}({\bf r})=N_{\da}.
\ee 

\begin{figure}[htb]
\begin{center}
\includegraphics[height=5cm]{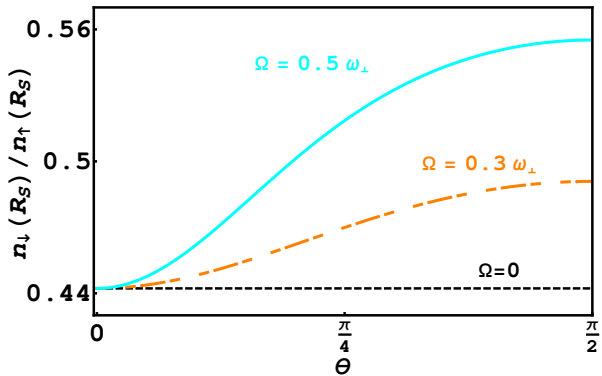}
\end{center}
\caption{(Color online) Concentration $n_\da/n_\ua$ for $P=44\%$, $\Om=0$ (black small dashed), $\Om=0.3\op$ (orange dashed), and $\Om=0.5\op$ (turquoise solid) as a function of the polar angle $\theta$.}
\label{fig:conzP44}
\end{figure}

\begin{figure}[htb]
\begin{center}
\includegraphics[height=5cm]{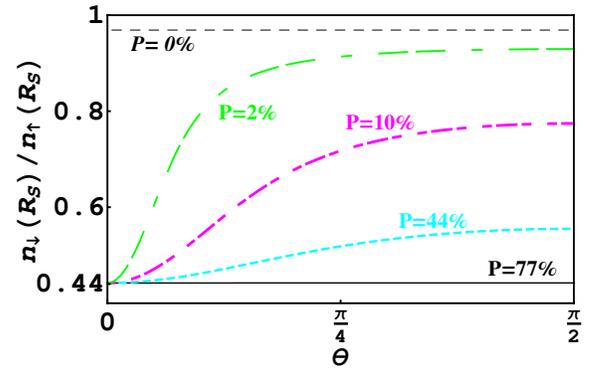}
\end{center}
\caption{(Color online) Concentration $n_\da/n_\ua$ at the superfluid-normal interface as a function of the polar angle $\theta$ for $\Om=0.5\op$ and different values of the polarization: $P=77\%$ (black solid), $P=44\%$ (turquoise dashed), $P=10\%$ (pink dotted-dashed), $P=2\%$ (green long dashed) and $P=0$ (black thin dashed).}
\label{fig:conPOL}
\end{figure}

In Fig. \ref{fig:conzP44} we plot the concentration $n_\da/n_\ua$ at the interface for $P=44\%$ as a function of the polar angle $\theta$ for different values of the angular velocity. The figure clearly points out the increase of the concentration when one moves from the $z$ axis to the $xy$ plane. 

Complementary to this is Fig. \ref{fig:conPOL}, where we show the concentration $n_\da/n_\ua$ for a fixed angular velocity $\Om=0.5\op$ and different values of the polarization as  a function of the angular position. This clearly reveals the nature of the normal state by highlighting the two extreme and singular cases $P=77\%$ and $P=0\%$. As already pointed out, at the threshold value $P=77\%$ for the nucleation of the superfluid, the rotation does not affect the value of $\xc$ as evidenced by the solid black line in Fig. \ref{fig:conPOL}. On the other hand, in the case that $P=0\%$ the critical concentration is constant and singular $n_\da/n_\ua=1$ for all angles but $\theta=0$. For all other values $0<P<\pc$, the rotation has a considerable impact on the local value $n_\da/n_\ua$.

\begin{figure}[htb]
\begin{center}
\includegraphics[height=5cm] {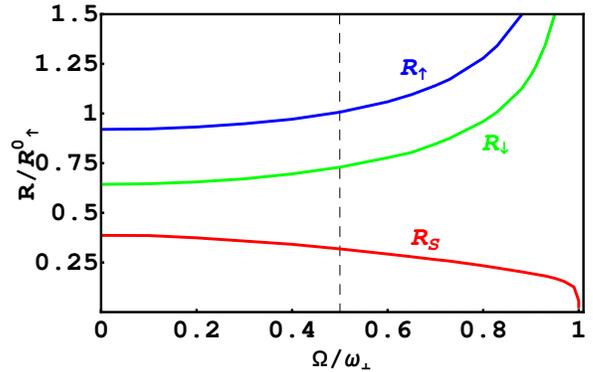}
\end{center}
\caption{(Color online) Radii in units of $R_{\ua}^0$ versus angular velocity for $\theta=\pi/2$ of the superfluid (red), $\uden$ (blue), and $\dden$ (green) for a polarization $P=44\%$.}
\label{fig:radiiP44}
\end{figure}

The density profiles  exhibit a typical shell structure.
In Fig. \ref{fig:radiiP44} we plot the  radii of the superfluid (red), $\ua$ (blue), and $\da$ (green) component in units of the Thomas-Fermi radius of an ideal gas $R_{\ua}^0$ versus the angular velocity $\Om/\op$ for a polarization $P=44\%$. While the superfluid radius decreases until the superfluid core completely vanishes at $\Om=\op$, the Thomas-Fermi radii of the $\ua$ and $\da$ component diverge for $\Om=\op$ as a consequence of the centrifugal effect. It is curious to see that while the normal part exhibits the typical bulge effect, the superfluid behaves in the opposite way. In fact, its radial size becomes  smaller than the axial one as a consequence of the depletion caused by the rotation, with  consequently inversion of the behavior of the aspect ratio $R_\perp/R_Z$ (see Fig. \ref{fig:sketch}).

\begin{figure}[htb]
\begin{center}
\includegraphics[height=5cm] {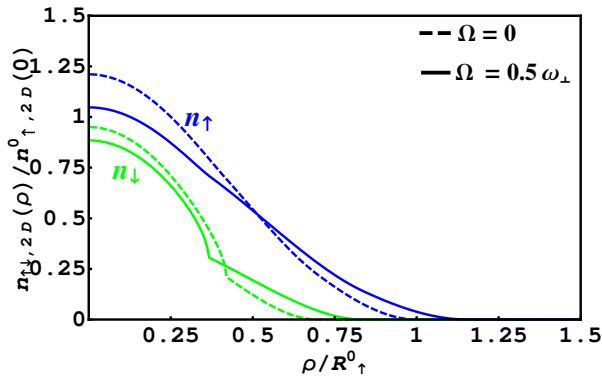}
\end{center}
\caption{(Color online) Column densities of the $\uden$ (blue),   and $\dden$ (green) component in a spherical harmonic trap for a polarization $P=44\%$ and $\Om=0$ (dashed lines) and $\Om=0.5\op$ (solid lines).}
\label{fig:CDtogether}
\end{figure}

\begin{figure}[htb]
\begin{center}
\includegraphics[height=5cm] {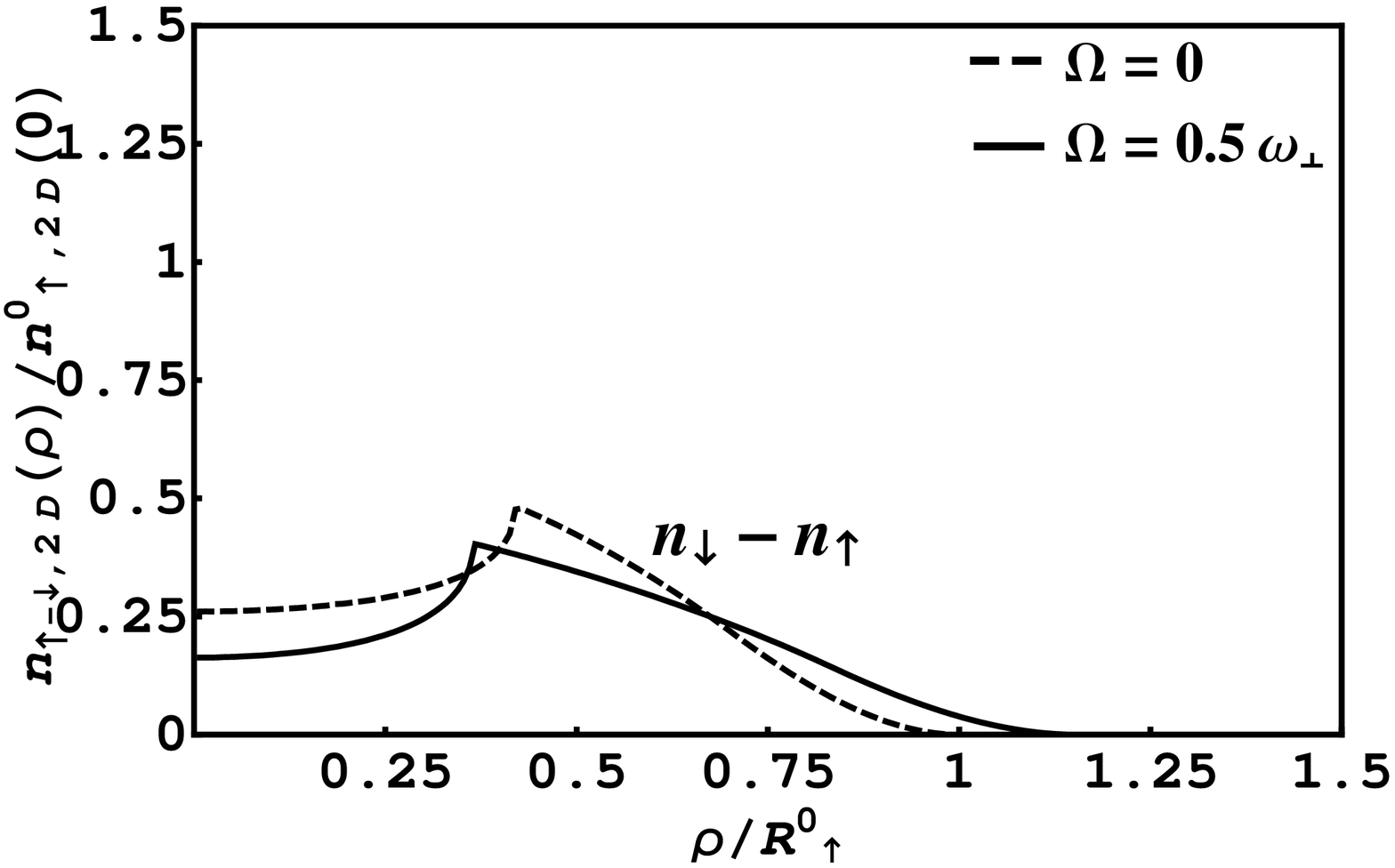}
\end{center}
\caption{Difference of the column densities of the $\uden$ and $\dden$ components in a spherical harmonic trap for a polarization $P=44\%$ and $\Om=0$ (dashed line) and $\Om=0.5\op$ (solid line).}
\label{fig:CDdiff}
\end{figure}

\begin{figure}[htb]
\begin{center}
\includegraphics[height=5cm] {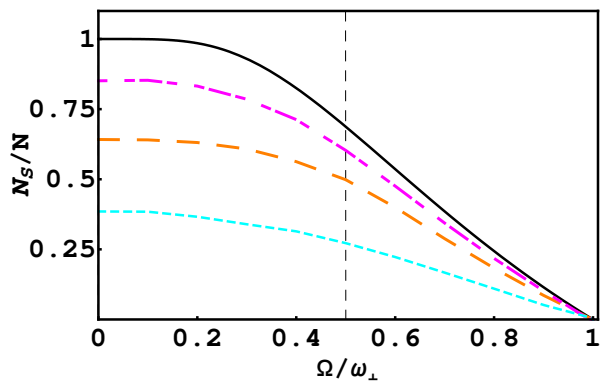}
\end{center}
\caption{(Color online) Evolution of the superfluid particle number for different polarizations  $P=0\%$ (black solid ), $P=10\%$ (pink dot-dashed) , $P=25\%$ (orange large dashed) and $P=44\%$ (turquoise small dashed) as a function of $\Om/\op$.}
\label{fig:NS}
\end{figure}

\begin{figure}[htb]
\begin{center}
\includegraphics[height=5cm] {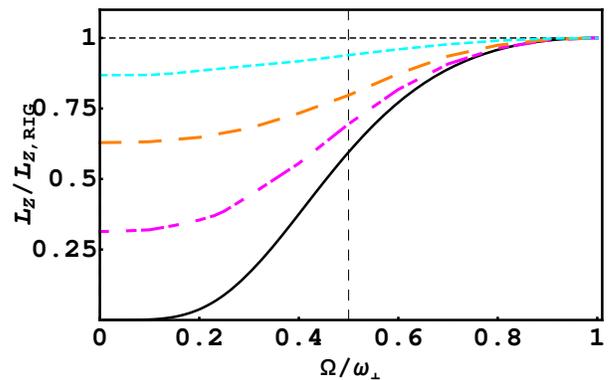}
\end{center}
\caption{(Color online) Angular momentum $L_{\rm Z}$ in units of the rigid value for different polarizations (black solid $P=0$, pink dot-dashed $P=10\%$, orange large dashed $P=25\%$ and turquoise small dashed $P=44\%$) as a function of $\Om/\op$.}
\label{fig:AM}
\end{figure}

It is worth mentioning that at large enough angular velocities the system exhibits solutions which break the axial symmetry \cite{alessio}. Such critical value is predicted to be $\Om_{\rm {cr}}\sim0.5\op$ as we discuss in detail in Sec. \ref{symmbreak}. 
The results in Figs.\ref{fig:radiiP44}, \ref{fig:NS}, and \ref{fig:AM} for $\Omega>0.5\op$ (dashed vertical line) correspond to the axial symmetric solution of the problem.

\begin{figure}[htb]
\begin{center}
\includegraphics[height=5cm] {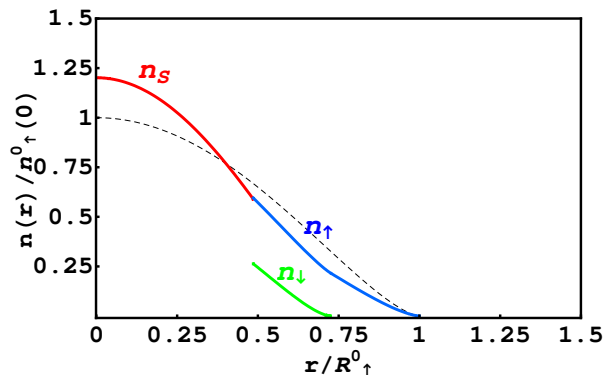}
\end{center}
\caption{(Color online) Density profiles for $\theta=0$ of the superfluid (red), $\uden$ (blue), and $\dden$ (green) in a spherical harmonic trap for a polarization $P=44\%$ and $\Om=0.5\op$ in units of the central density of the noninteracting gas (dashed line).}
\label{fig:dens0}
\end{figure}

\begin{figure}[htb]
\begin{center}
\includegraphics[height=5cm] {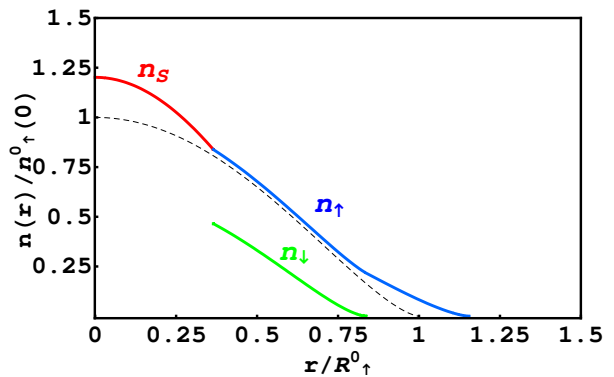}
\end{center}
\caption{(Color online) Density profiles for $\theta=\pi/2$ of the superfluid (red), $\uden$ (blue), and $\dden$ (green) in a spherical harmonic trap for a polarization $P=44\%$ and $\Om=0.5\op$ in units of the central density of the noninteracting gas (dashed line).}
\label{fig:dens06}
\end{figure}

\begin{figure}[htb]
\begin{center}
\includegraphics[height=5cm] {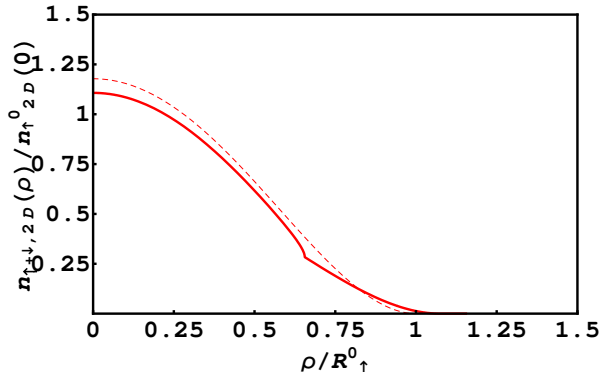}
\end{center}
\caption{(Color online) Total column density of the unpolarized system in a spherical harmonic trap for $\Om=0$ (dashed red line, superfluid at rest). For $\Om=0.5\op$ the system consists of a superfluid core surrounded by a normal shell where $\uden=\dden$.}
\label{fig:CDsf}
\end{figure}

In an experiment the effect of phase separation as well as the radius of the superfluid are best revealed as a knee in the {\sl in situ} column density $n_{\sigma,2D}(\rho)\equiv\int dz\  n_{\sigma}({\bf r})$, with $\sigma=\ua,\da$. These observables can nowadays be measured with high precision using phase-contrast image techniques. We expect that the position of the knee for a fixed polarization will depend on the  angular velocity. This is clearly shown in  the column density of the majority ($\ua$, blue) and the minority ($\da$, green) components, Fig. \ref{fig:CDtogether}, for $\Omega=0$ (dashed) and $\Om=0.5\op$ (solid), as well as in the density difference, Fig. \ref{fig:CDdiff}.
The knee is a direct consequence of  the discontinuity exhibited by the three dimensional density shown in Figs.\ref{fig:dens0} and \ref{fig:dens06}, where we plot  $n_{\rms}$ (red), $n_{\ua}$ (blue), and $n_{\da}$ (green) in a spherical trap for $\theta=0$ and $\theta=\pi/2$, respectively. The densities and the radial coordinate have been renormalized with respect to the central value of $n_{\ua}^0$ and the Thomas-Fermi radius $R_{\ua}^0$ of an ideal gas. In accordance to the results for the superfluid radius shown in Fig. \ref{fig:radiiP44} the discontinuity in the density takes place at a smaller value of the radius compared to the nonrotating configuration.
It is worth noticing that this knee is also exhibited by the total density of the unpolarized gas (Fig. \ref{fig:CDsf}), reflecting the density discontinuity produced by the rotation \cite{ias}.

For $\theta=0$ the densities $n_{\ua}$ and $n_{\da}$ jump from the superfluid value $n_{\rms}$ to the values $n_{\ua}\sim 1.01n_{\rms}$ and $n_{\da}=\xc  n_{\ua}\simeq0.44n_{\rms}$ as one enters the normal phase, precisely as in the nonrotating case \cite{normalLobo}.
Yet for $\theta=\pi/2$ the behavior is different and, in particular, for $\Om=0.5\op$ the densities jump from $n_{\rms}$ to  $n_{\ua}\simeq 0.99n_{\rms}$ and $n_{\da}\simeq0.55n_{\rms}$. The smaller relative jump with respect to the nonrotating case reflects the smaller polarization (higher concentration $x$) exhibited by the rotating normal phase at the interface.

Further insight on the effect of the rotation is provided by the depletion of the superfluid. In Fig. \ref{fig:NS} we plot the ratio between the number of particles in the superfluid $N_{\rm S}$ and the total number $N$ (superfluid fraction) as a function of the angular velocity. This effect is especially pronounced for small polarizations (Fig. {\ref{fig:NS}} black line, $P=0$) since the depletion for higher polarization is large already in the nonrotating case.  

Finally in Fig. \ref{fig:AM} we plot the angular momentum 
$L_{\rm Z}=\Om\int d{\bf r} (x^2+y^2) n_{\rm N}({\bf r})$ of the system for different polarizations. For an axisymmetric configuration the superfluid does not carry angular momentum which is then provided only by the normal component.  Hence, the more particles are in the normal part, the stronger the response of the system to the rotation.

\section{Quadrupole Instability induced by the rotation}
\label{symmbreak}

In the preceding section we have shown that the effect of the rotation applied to a polarized Fermi gas at  unitarity is to enhance the depletion of the superfluid.

A further important effect is the spontaneous breaking of axi-symmetry. In fact, one expects that above a certain critical value $\Om_{\rm cr}$ the system exhibits a surface energetic instability, undergoing  a continuous shape deformation, similarly to what happens in Bose-Einstein condensates \cite{alessio,Dalibard1,Dalfovo}. This effect is accounted for by the solution ${\bf v_{\rm S}} \ne{\bf 0}$ of  Eq.(\ref{eq:graddens}) which becomes energetically favorable above $\Om_{\rm cr}$.  

In the case of Bose-Einstein condensates the quadrupole instability occurs at $\Omega_{\rm{cr}}= \omega_\perp/\sqrt 2$ when the $\omega=\sqrt{2}\omega_\perp$ hydrodynamic quadrupole mode  becomes energetically unstable in the rotating frame \cite{alessio}. 
A different value  is  predicted to occur for the rotating Fermi gas, due to the new boundary conditions imposed by the presence of the normal component \cite{note4}. In fact,  the current of the superfluid, in the rotating frame, should be  tangential to the interface where the density, differently from the Bose-Einstein condensate (BEC) case, does not vanish. 

In the following we will determine the value of the critical velocity in the simplest case of a rotating unpolarized gas and we will consider the onset of a quadrupole deformation. 

We consider an axially symmetric potential $\ox=\oy=\op\neq\oz$  and a solution where axi-symmetry is spontaneously broken. This is associated with the appearance of a non vanishing velocity field in the superfluid component whose velocity potential will be chosen of the form 
\be
\Phi=xy f(r^2,z^2),
\label{eq:velfield}
\ee
where $f$ is an arbitrary function of $r^2=x^2+y^2$ and of $z^2$. Note that in the case of the quadrupole instability of a Bose-Einstein condensate an exact stationary solution is found with $f=$const.

The value of $\Omega_{\rm cr}$ corresponds to the onset of solutions with a deformed configuration. It is determined by solving the continuity equation in the rotating frame
\be
\nabla \cdot[(\nabla \Phi -{\bf \Omega} \times {\bf r})n_{\rm S}({\bf r})]=0,
\label{eq:conteq}
\ee
where $n_{\rm S}({\bf r})$ is now no longer axisymmetric and by imposing that  
the superfluid current be tangential to the boundary $B(\bf{r})$ of the superfluid
\be
\left.(\nabla\Phi-{\bf{\Om\times r}})\cdot\nabla B({\bf{r}})\frac{}{}\right|_{r=\suf}=0.
\label{eq:TC}
\ee

The density, in the local density  approximation,  is given by 
\be
n_{\rms}({\bf r})=\frac{1}{\xi_{\rms}}\frac{1}{6\pi^2}\left(\frac{2m}{\hbar^2}\right)^{3/2}(\mu-V^{\rms}({\bf{r}}))^{3/2},
\ee
where $V^{\rm S}({\bf r})=V({\bf r})+\frac{1}{2}mv_{\rm S}^2-m\Omega({\bf r}\times{\bf v}_{\rm S})_{\rm Z}$ is the effective harmonic potential felt by the superfluid. By expanding the external potential to first order in $f$
as $V^{\rm S}({\bf r})=V({\bf r})+\delta V^{\rm S}({\bf r})$ with  $\delta V^{\rm S}({\bf r})=-m\nabla\Phi(\bf{\Om\times r})$, one finds $n_{\rms}({\bf r})=(\mu-V({\bf r}))^{3/2}-\frac{3}{2}(\mu-V({\bf r}))^{1/2}\cdot\delta V^{\rms}({\bf r})$.
At the same time the border can be written as  $B({\bf{r}})=B^0({\bf{r}})+\delta B({\bf{r}})$, where $B^0(\bf{r})$ is the radius of the superfluid  given by \cite{ias}   
\be
R_{\rms}^2(\theta)=\frac{2\mu}{m}\left(\oz^2\cos^2\theta+\op^2\sin^2\theta+\frac{\gamma}{1-\gamma}\Om^2\sin^2\theta\right)^{-1},
\label{eq:rad}
\ee
and $\delta B({\bf{r}})=-m\nabla\Phi\cdot(\bf{\Om\times r})$ is the linear perturbation due to the quadrupole symmetry breaking.

Then, Eqs.(\ref{eq:conteq}) and (\ref{eq:TC}) reduce to 
\be
(\mu-V({\bf r}))\Delta\Phi+\frac{3}{2}(\nabla\delta V^{\rms}({\bf r}))\cdot({\bf\Om\times r})-\frac{3}{2}\nabla V({\bf r})\cdot\nabla\Phi=0,
\label{eq:lincont}
\ee
and 
\be
\left.\nabla\Phi\cdot\nabla B_0({\bf r})-({\bf \Om\times r})\cdot\nabla\delta B({\bf r})\frac{}{}\right|_{r=\suf(\theta)}=0.
\label{eq:linTC}
\ee
By inserting the explicit expressions of the respective functions in Eqs.(\ref{eq:lincont}) and (\ref{eq:linTC}), after some straightforward  algebra \cite{Ingridthesis} we obtain 
\beq
& &(2\Om^2-1)f+\frac{2}{3}(1-r^2-z^2)(r^2f_{\rm rr}+z^2f_{\rm zz})\nonumber\\
& &+(2-3r^2-2z^2)f_{\rm r}+\frac{1}{3}(1-r^2-4z^2)f_{\rm z}=0,
\label{eq:contF}
\enq
and 
\beq
&&\left[\frac{}{}(1-\gamma(1-\Om^2)-2\Om^2)f\right.+(1-\gamma(1-\Om^2))r^2f_{\rm r}\nonumber\\
&&+\left.(1-\gamma)z^2f_{\rm z}\frac{}{}\right](1-r^2-z^2)^{3/2}\left.\frac{}{}\right|_{r=\suf(\theta)}=0,
\label{eq:TCF}
\enq
where the latter equation is evaluated at the interface Eq.(\ref{eq:rad}).
In Eqs.(\ref{eq:contF}, \ref{eq:TCF}) $f_{\rm r(\rm z)}$ is the first derivative of $f$ with respect to $r$ ($z$), $f_{\rm rr(\rm zz)}$ the second, and the expressions have been renormalized with respect to the radius of a superfluid at rest $(R_{\rms}^0)^2=2\mu/m\op^2$ and we have made the substitution $\Om/\op\rightarrow\Om$ and the assumption $\oz=\op$. 

In the case of a two-dimensional  system ($\suf(\theta)\equiv\suf$) the previous equations can be easily solved. In this case indeed Eqs.(\ref{eq:contF}) and (\ref{eq:TCF}) reduce to 
\be
(2\Om^2-1)f(r^2)+(2-3r^2)f_{\rm r}(r^2)+\frac{2}{3}(1-r^2)r^2f_{\rm rr}(r^2)=0,
\label{eq:contF2D}
\ee
and 
\beq
&&[1-\gamma(1-\Om^2)-2\Om^2]f(r^2)\nonumber\\
&+&[1-\gamma(1-\Om^2)]r^2f_{\rm r}(r^2)\left.\frac{}{}\right|_{r=\suf}=0,
\label{eq:TCF2D}
\enq
\newline
the latter being evaluated at
\be
\suf^2\equiv \frac{(1-\gamma)}{[1-\gamma(1-\Om^2)]} .
\label{eq:normquad}
\ee

The solutions of the continuity equation Eq.(\ref{eq:contF2D})  are hypergeometric functions $_{2}F_{1}(a,b,c;r^2)$ \cite{stegun}, with the coefficients $a,b$ and $c$ given by 
\beq
a&=&\frac{7}{4}-\frac{1}{4}\sqrt{25+48\Om^2},\nonumber\\
b&=&\frac{7}{4}+\frac{1}{4}\sqrt{25+48\Om^2},\nonumber\\
c&=&3.
\label{eq:coeff2D}
\enq

Inserting these solutions in the boundary condition (\ref{eq:TCF2D}), we find the value $\Omega_{\rm cr}=0.45 \omega_{\perp}$ for the emergence of a spontaneous quadrupole deformation in two dimensions. 

For the three-dimensional case the calculation is more cumbersome and can be solved only numerically. 
The results of the numerical calculation yield the estimate $\Omega_{\rm cr}\sim 0.5\omega_{\perp}$ for the critical angular velocity in three dimensions. Notice that both in two and  three dimensions the critical velocity is predicted to be smaller than the value $\Omega_{\rm cr}= \omega_{\perp}/\sqrt{2}$ holding in the BEC case.

\section{Conclusions}
\label{con}

In conclusion we have analyzed the effect of adiabatic rotation on a polarized Fermi gas at unitarity assuming phase separation between a superfluid and a normal phase. 
We find that the normal phase is energetically favoured by the rotation and thus the superfluid is further depleted with
respect to the nonrotating configuration.
The normal region exhibits the typical bulge effect due to the centrifugal force while the superfluid is squeezed. This has clear observable effects on the density profiles which can be addressed experimentally. 
A striking feature is that although the {\sl global} polarization is not affected by the rotation, the concentration $n_{\da}/n_{\ua}$ at the border increases from the non-rotating value on the $z$ axis to a maximum value in the $xy$ plane. 

We have also addressed the question of quadrupole instability of the superfluid core, which produces a spontaneous breaking of axial symmetry of the cloud. 
The critical frequency for the onset of the instability turns out to be smaller than in the BEC case.
Its measurement would provide a further crucial test of the mechanism of phase separation and of the equation of state  of the normal phase \cite{normalLobo,recati,shinalone}.

In our work we assume that the polarized system phase separates in only two phases. This assumption works extremely well for the experiment carried out so far \cite{recati}. In the rotating case other phases could show up and it would be very interesting to see how they affect our
results. For example, within BCS theory a third superfluid phase is found to occupy a small region at the interface \cite{urban}. Unfortunately we know that BCS theory is quantitatively (and sometimes also qualitatively) not correct. A more microscopic investigation of the phase separation at unitarity would be neccessary to settle the problem.

Finally, let us briefly argue on the possibility of having a rotating system without vortices in the superfluid, which is the main assumption of the
present work. If the vortices could enter the superfluid, the lowest energy configuration would not be the one discussed in the paper, but rather a superfluid core with vortices sorrounded by a rotating normal shell, as also experimentally seen in \cite{MIT1}. We expect that there exists a
barrier for a vortex to enter the superfluid, as happens for a BE condensate. The interface caused by the polarized fermions could however
change the scenario. On the one hand, the phase separation in the trap should disfavor the vortex formation as the superfluid density is finite at the interface. On the other hand, since there is a relative velocity between the superfluid and the normal shell, vortex nucleation could be favoured by a Kelvin-Helmoltz-like mechanism. Moreover, we have shown that the interface reduces the critical angular velocity for a quadrupole instability. In the BEC case the latter is considered as a route toward vortex formation \cite{Dalibard1}. In the end, more theoretical and experimental work is needed to enlighten the issue of vortex nucleation in these polarized systems.

\section{acknowledgments}

We acknowledge very useful discussions with Frederic Chevy, Stefano Giorgini, Carlos Lobo, Lev P. Pitaevskii, Christophe Salomon, Alberto Valli and Martin Zwierlein. We also acknowledge support by the Ministero dell'Istruzione, dell'Universit\`a e della Ricerca (MIUR) and by the EuroQUAM FerMix program.


\begin{thebibliography}{100}
\bibitem{noteUrban}
In the rotating case a third phase occupying a very small region between the superfluid and the normal one 
has been recently identified in \cite{urban} using a BCS mean-field approach.
\end{thebibliography}
\end{document}